\documentclass[showpacs,twocolumn,pre,floatfix]{revtex4}
\usepackage{psfrag,epsfig,amsfonts,amssymb,amsmath,wasysym,bm}
\usepackage{dcolumn}
\usepackage{mathtools}

\usepackage[normalem]{ulem}
\usepackage{color}

\newcommand{\D}{q}
\newcommand{\QQ}{Q}
\newcommand{\KK}{K}
\newcommand{\kk}{k}
\newcommand{\Do}{q_{opt}(\rho(t),\bar\rho,\vec s)}
\newcommand{\slp}{\mbox{max}'_n p_n}

\newcommand{\SP}{P^{succ}}
\newcommand{\ord}{{\cal O}}
\newcommand{\propa}{{\cal U}_t}

\newcommand{\tr}{\mbox{Tr}}
\newcommand{\da}{\Delta_{\! A}}

\newcommand{\NN}{{\mathbb N}}

\begin{document}

\title{Equilibration of isolated many-body 
quantum systems with respect to general 
distinguishability measures}

\author{Ben N. Balz}
\author{Peter Reimann}
\affiliation{Universit\"at Bielefeld, Fakult\"at f\"ur Physik, 
33615 Bielefeld, Germany}

\begin{abstract}
We demonstrate equilibration of isolated many-body 
%quantum 
systems in the sense that, after initial 
transients have died out, the system behaves 
practically indistinguishable from a time-independent 
steady state, i.e., non-negligible deviations are 
unimaginably rare in time.
Measuring the distinguishability in terms of
quantum mechanical expectation values,
%Taking as distinguishability measure the
%expectation values of quantum mechanical observables, 
results of this type have been
previously established under increasingly 
weak assumptions about the initial disequilibrium, 
the many-body Hamiltonian, and the 
considered observables.
Here, we further extend these results with
respect to generalized distinguishability 
measures which fully take into account 
the fact that the actually observed,
primary data are not expectation 
values but rather the probabilistic 
occurrence of different possible 
measurement outcomes.
\end{abstract}


\maketitle

%%%%%%%%%%%%%%%%%%%%%%%%%%%%%%%%%%%%%%%%%%%%%%%%
\section{Introduction}
\label{s1}
Does a macroscopic system, prepared in a 
non-equilibrium initial state, and evolving in 
isolation from the rest of the world,
approach a steady state in the long time limit?
Due to quantum revivals, time inversion invariance,
and other quite obvious reasons (see Sect. \ref{s3a}), 
such a relaxation towards an equilibrium state 
can certainly not be true in the strict sense.
On the other hand, ``practical equilibration''
has been established in 
Refs. \cite{equil,sho11,sho12,rei12}
under quite weak conditions in the sense 
that the expectation value of quantum 
mechanical observables remains extremely 
close to a constant value for the overwhelming 
majority of all sufficiently late times.
In other words, deviations of expectation 
values from a steady long time limit are 
either so small or so rare that they can 
be safely neglected in any real experiment.

Yet, it has been pointed out by Short 
\cite{sho11} that these findings are still not 
fully satisfactory since the primary data
in a quantum mechanical measurement
are not expectation values but rather the 
%apparently random 
probabilistic
occurrence of different possible 
measurement outcomes.
%, rather than the expectation
%value of an observable.
Indeed, the exact probabilities of 
those outcomes are generically 
not strictly identical to the presumed 
steady state values,
hence the difference 
unavoidably must 
become statistically resolvable when
repeating the measurement sufficiently 
many times.
If such a difference would already be 
recognizable by an experimentally 
feasible
number of repetitions, 
then practical equilibration in the above 
sense would thus no longer hold true.
The main purpose of our present paper is
to exclude the latter possibility by
further developing the approach 
from Refs. \cite{sho11,sho12,rei12}.

Two immediate next questions are:
Does the steady long time limit 
agree (at least approximately) 
with the value predicted by
one of the canonical ensembles
from textbook statistical physics?
What is the characteristic time 
scale governing the relaxation 
towards equilibrium?
Both questions are clearly of great
conceptual as well as practical
importance, and they have been recently
addressed by numerous analytical,  
numerical, and experimental
works, see Refs. \cite{neu29,eth,times,exp} 
for a few representative examples.
Yet, these questions are beyond 
the scope of our present paper.

%%%%%%%%%%%%%%%%%%%%%%%%%%%%%%%%%%%%%%%%%%%%%%%%
\section{General framework}
\label{s2}
We focus on time-independent Hamiltonians 
of the form
\begin{equation}
H = \sum_n E_n P_n \ ,
\label{1}
\end{equation}
where the $P_n$ are projectors onto the 
eigenspaces of $H$ with mutually different
eigenvalues $E_n$,
and where $n$ runs from $1$ to infinity or to 
some finite upper limit.

The main examples we have in 
mind are isolated macroscopic 
systems with, say, $f\approx 10^{23}$ 
degrees of freedom.
Compound systems, consisting of a
subsystem of actual interest
and a much larger environmental bath, 
are thus included as special cases.
While the precise requirements on $H$ 
will be provided later, we anticipate 
that, similarly as in Refs. 
\cite{equil,sho11,sho12,rei12},
those rather weak requirements
do {\em not} imply that the system must 
be ``non-integrable'' or ``chaotic'' 
in the sense of Refs. \cite{eth}.
%still admit many so-called 
%integrable or non-chaotic 
%systems \cite{eth}.
Moreover, our explorations may also 
be of interest, e.g., for systems 
with few degrees of freedom \cite{xx}
but with a high dimensional ``active
Hilbert space'' 
\footnote{Here ``a high dimensional 
active Hilbert space'' means that
-- analogously as in 
Eqs. (\ref{7}), (\ref{8}) --
the energy level populations $p_n$
must be small for all but one $n$.
Apart from trivial cases this implies 
that there must be many levels which 
are non-negligibly populated by the 
system state (``active''), and which 
thus span a high dimensional Hilbert 
space.}.
%The span of the corresponding 
%eigenvectors is the so-called
%``active Hilbert space''.}.

As usual, system states (pure or mixed) 
are described by density operators 
$\rho$ and
observables by Hermitian
operators $A$.
Expectation values are given by 
$\langle A\rangle_{\!\rho}:= \tr\{\rho A\}$
and the time evolution by
$\rho(t)=\propa\rho(0)\propa^\dagger$
with propagator $\propa:=e^{-iHt}$
($\hbar=1$),
implying with (\ref{1}) that
$\propa:=\sum_n P_n e^{-iE_nt}$ 
and hence that
\begin{eqnarray}
%\tr\{\rho(t) A\}
\langle A\rangle_{\!\rho(t)}
& = & \sum_{m,n}
A_{mn}
\,
 e^{i (E_n-E_m) t}
\label{2}
\\
A_{mn}& := &\tr\{P_m \rho(0) P_n A \}
\ . 
\label{3}
\end{eqnarray}

Similarly as in (\ref{1}), any given 
observable $A$ can be written in 
the form
\begin{equation}
A=\sum_{\nu} a_{\nu}\, \KK_{\nu} \ ,
\label{4}
\end{equation}
where $\KK_{\nu}$ are the projectors 
onto the eigenspaces of $A$ 
with mutually different eigenvalues 
$a_{\nu}$.
According to textbook quantum mechanics,
%use the textbook approach that
%quantum mechanical measurememt theory:
%description of the measurement process:
%For the sake of simplicity only,
%we furthermore restrict ourselves 
%to the simplest and most common 
%approach 
given a system in state $\rho(t)$, any single 
%quantum mechanical 
measurement 
%process 
of the observable $A$
results in one of the possible 
outcomes $a_{\nu}$, and the probability 
to obtain the specific 
outcome $a_{\nu}$ is given by
\begin{equation}
\kk_{\nu}(t):=\tr\{\rho(t)\KK_{\nu}\} \ .
\label{5}
\end{equation}
As an aside, we remark that our present 
approach could be readily extended to 
%the description of quantum mechanical measurements
%%process 
%in terms of
the so-called positive-operator 
valued measure (POVM) formalism
\cite{povm}:
The only formal difference 
would be that the pertinent 
operators $K_\nu$ are then 
in general no longer mutually orthogonal, 
a property which we never 
actually exploit in the following.

Finally, we adopt the viewpoint that
no experimentally realistic measurement
yields more than, say,  $20$ relevant digits.
Hence, 
%we can restrict ourselves to 
it is sufficient to consider observables
%$A$ 
with less than $10^{20}$ different 
measurement outcomes $a_\nu$,
i.e., we can and will restrict ourselves 
from now on to observables $A$ which 
satisfy the conditions
\begin{eqnarray}
\nu\in\{1,2,...,N_A\}\ , \ N_A < 10^{20} \ .
\label{6}
\end{eqnarray}

In view of Eq. (\ref{1}), the specific 
observable $A=P_n$ describes
the population of the (possibly degenerate)
energy level $E_n$ with expectation value 
(occupation probability)
\begin{equation}
p_n:=\tr\{ \rho(t) P_n \} \ .
\label{7}
\end{equation}
Since $P_n$ commutes with 
$H$ from (\ref{1}), 
the level populations $p_n$ 
are $t$-independent 
(conserved quantities).
Thus they are entirely determined 
already by the initial condition
(system preparation).

For typical macroscopic systems with 
$f\approx 10^{23}$ degrees of freedom,
the energy levels are unimaginably dense.
Under realistic experimental conditions
it is therefore practically unavoidable to notably
populate a number of energy levels 
which is exponentially large in $f$.
In turn, every single level population 
$p_n$ from (\ref{7}) is expected to 
be extremely 
small (compared to $\sum_n p_n=1$)
and to typically satisfy the 
very rough estimate 
$\max_n\, p_n=10^{-\ord(f)}$.
In the following, we even admit
the possibility that one single 
energy level exhibits a non-small
population, for instance a 
macroscopically populated 
ground state.
Accordingly, we may still expect that
\begin{equation}
{\max_n}' p_n=10^{-\ord (f)} \ ,
\label{8}
\end{equation}
where $\slp$ indicates 
the second largest energy 
level population.

%%%%%%%%%%%%%%%%%%%%%%%%%%%%%%%%%%%%%%%%%%%%%%%%
\section{The problem of equilibration}
\label{s3}
%%%%%%%%%%%%%%%%%%%%%%%%%%%%%%%%%%%%%%%%%%%%%%%%
\subsection{Preliminary results}
\label{s3a}
Our preliminary formulation of the problem of
equilibration consists in the question whether, 
in which sense, and under what conditions the 
expectation value from (\ref{2}) approaches 
a constant 
%``steady state'' 
value in the long time limit.

It is quite obvious that the expectation 
value (\ref{2}) cannot rigorously converge 
towards any long time limit apart from trivial 
cases with $A_{mn}=0$ for all $m\not =n$.
%(the latter follows, e.g. if $A$ is a 
%conserved quantity or if $\rho(t)$ is time 
%independent right from the beginning).
Moreover, it is well known that any $\rho(t)$
returns arbitrarily close to the initial 
state $\rho(0)$ for certain, sufficiently 
late time points $t$ (quantum revivals).

The only remaining hope is that 
(\ref{2}) {\em approximately} approaches
some steady value
% $\langle A\rangle_{\!\bar \rho}$
for {\em most} sufficiently large times $t$.
Intuitively, if any such steady 
asymptotics is approached at all, then the 
most promising candidate appears to be
the value which is obtained by averaging 
(\ref{2}) over all times $t\geq 0$.
Since all energies $E_n$ are mutually 
different (see below Eq. (\ref{1})), 
one readily can infer from (\ref{2}) 
that this putative steady state 
should thus be given by 
$\langle A\rangle_{\!\bar \rho}:= 
\tr\{\bar\rho A\}$, 
where
\begin{eqnarray}
\bar \rho := \sum_n P_n\rho(0)P_n 
\label{9}
\end{eqnarray}
is a non-negative Hermitian 
operator of unit trace and thus a 
well defined density operator.
A result of this type is derived in 
Appendix A by 
combining and refining techniques
% combining and simplifying arguments 
originally due to 
Refs. \cite{sho12,rei12,het15}.
Namely, it is shown that the following
inequality holds for all sufficiently 
large $T$:
\begin{eqnarray}
\frac{1}{T}\int_0^T  dt\, [\sigma (t)]^2 
& \leq & 
%24 
3 \, \tr\{\bar\rho A^2\}  \, g\, {\max_n}' p_n
\label{10}
\\
\sigma(t) & := & 
\langle A\rangle_{\!\rho(t)} - \langle A\rangle_{\!\bar\rho}
\ ,
\label{11}
\end{eqnarray}
%where $\da$ is the range of $A$,
%i.e., the difference between its 
%largest and smallest eigenvalues 
%(cf. (\ref{4}) and (\ref{6})), and 
where $\slp$ is the second largest 
energy level population (see around Eq. (\ref{8})) 
and $g$ represents the maximal 
degeneracy of energy gaps,
\begin{equation}
g := 
\max_{m\not = n}|\{(k,l) \, : \, E_k-E_l=E_m-E_n\}|  \ ,
\label{12}
\end{equation}
with $|S|$ denoting the number of elements
contained in the set $S$.
In other words, $g$ is the maximal number of 
exactly coinciding energy differences among 
all possible pairs of distinct energy eigenvalues.

In view of (\ref{8}) and disregarding exceedingly
large gap degeneracies $g$, the time average
on the left hand side in (\ref{10}) is extremely 
small, implying that the deviation (\ref{11}) must
be very small in modulus for most times 
$t \in [0,T]$.
In order to quantify this argument, we define for 
any given $\epsilon>0$ and $T>0$ the quantity
\begin{eqnarray}
T_{\epsilon} := 
\big| 
\{\, 0 \leq t  \leq T\, : \, |\sigma(t)|  > \epsilon\, \}
\big| \ ,
\label{13}
\end{eqnarray}
where $|S|$ denotes the size (Lebesgue measure) 
of the set $S$.
In other words, $T_{\epsilon}$ is the measure of all 
times $t\in [0, T]$ for which 
$|\sigma(t)| >  \epsilon$ holds true.
It follows that $[\sigma (t)]^2 > \epsilon^2$
for a set of times $t$ of measure 
$T_{\epsilon}$ and that
$0\leq [\sigma(t)]^2 \leq \epsilon^2$ for all 
remaining times $t$ contained in $[0, T]$.
Hence the temporal average on the left 
hand side of (\ref{10}) must 
be bounded from below by 
$\epsilon^2 T_{\epsilon}/T$.
It follows that for 
any given $\epsilon>0$
\begin{eqnarray}
T_{\epsilon}/T\leq  
%6\, (\da^2/\epsilon)^2 
%24
3 \, \tr\{\bar\rho A^2\}\, g\, {\max_n}' p_n/\epsilon^2
\label{14a}
\end{eqnarray}
for all sufficiently large $T$.

Note that the left hand side of (\ref{10}) 
remains unchanged if $A$ is replaced by 
$A+c\, 1$, where $c$ is an arbitrary real 
number and $1$ the identity operator. 
Accordingly, we may 
replace also on the right hand 
side of (\ref{10}) 
%and (\ref{14}) 
$A$ by $A+c\, 1$ 
with an arbitrary $c$.
Denoting by $a_{max}$ and $a_{min}$
the largest and smallest eigenvalues of $A$
(cf. (\ref{4}) and (\ref{6})), and by
$\da:=a_{max}-a_{min}$ the 
range of $A$,
we can and will choose $c$ so 
that $|a_\nu-c|\leq\da/2$ for all eigenvalues 
$a_\nu$ of $A$.
It follows that $\tr\{\bar\rho\, (A+c\,1)^2\}\leq (\da/2)^2$
on the right hand side of (\ref{10}),
and likewise in (\ref{14a}), i.e.
\begin{eqnarray}
\frac{T_{\epsilon}}{T}
\leq  
%6\, 
\frac{3\, \da^2\, g\, {\max_n}' p_n}{4\, \epsilon^2}
\label{14}
\end{eqnarray}
for all sufficiently large $T$.

Relation (\ref{14}) together with 
(\ref{8}), (\ref{12}),
and (\ref{13}) represents the
answer to the above stated, 
preliminary problem of equilibration:
For any given $\epsilon > 0$ the
``true'' expectation value 
$\langle A\rangle_{\!\rho(t)}$ deviates from 
the constant value $\langle A\rangle_{\!\bar\rho}$
by more than $\epsilon$ for a set of times
$t\in[0,T]$, whose measure $T_\epsilon$  
is bounded by (\ref{14}) for all 
sufficiently large $T$.
If $\epsilon$ as well as
the right hand side of (\ref{14}) 
are both sufficiently small,
which is easily feasible in view of (\ref{8}),
then it follows that $\langle A\rangle_{\!\rho(t)}$ 
is practically (within any experimentally 
achievable resolution) constant for the
overwhelming majority of all times 
$t\in[0,T]$.
In particular, $T$ must be so large 
that the initial decay process
(from the possibly far from equilibrium
initial value $\langle A\rangle_{\!\rho(0)}$
towards the equilibrium value 
$\langle A\rangle_{\!\bar\rho}$) 
is accomplished during a time interval 
much smaller than $[0,T]$.

Note that Hamiltonians with 
degenerate energy gaps are, loosely speaking, 
of measure zero among ``all'' Hamiltonians.
They only arise in the presence of special 
reasons like (perfect) symmetries, 
additional conserved quantities 
(besides $H$), or fine-tuning of 
parameters.
Generically, all non-trivial energy gaps 
$E_m-E_n$ (i.e., those with $m\not=n$) 
are thus mutually different, implying
$g=1$ in (\ref{12}).
Our above results remain valid even
%In full generality, the result 
%(\ref{10}) even remains valid 
for non-generic cases with $g > 1$.
Likewise, Hamiltonians with degenerate
energy eigenvalues are in principle
non-generic, but still admitted in (\ref{1}).

In summary, the true system state $\rho(t)$
becomes experimentally indistinguishable from 
the time independent approximation $\bar\rho$ 
for practically all sufficiently late times $t$ 
under very weak conditions on the
initial state, the observable, 
and the Hamiltonian.

%%%%%%%%%%%%%%%%%%%%%%%%%%%%%%%%%%%%%%%%%%%%%%%%
\subsection{Reformulation of the problem}
\label{s3b}
So far, the (non-)distinguishability of 
$\rho(t)$ and $\bar\rho$ was always 
meant with respect to the corresponding 
two expectation values of the 
considered observable $A$.
As pointed out by Short \cite{sho11},
such a distinguishability criterion 
%in terms of expectation values 
is not entirely satisfying since the 
basic measurable quantities are 
%actually {\em not} the 
{\em not}
expectation values, 
but rather the different possible 
measurement outcomes $a_\nu$, 
see above Eq. (\ref{5}).
Hence, the distinguishability
of $\rho(t)$ and $\bar\rho$ should 
be based on 
%a comparison between
the actually observed, random
occurrence of each outcome 
$a_\nu$.
More precisely, one should compare
in some suitable way the probabilities 
$\kk_\nu(t)$ from (\ref{5}) and
\begin{eqnarray}
%\kk_\nu & := & \tr\{\rho \KK_\nu\}
%\label{15}
%\\
\bar \kk_\nu& := & \tr\{\bar\rho \KK_\nu\}
\label{15}
\end{eqnarray} 
with which the different possible measurement
outcomes $a_\nu$ are observed in the two 
states $\rho(t)$ and $\bar\rho$, respectively.
Indeed, it could well be that
$\rho(t)$ and $\bar\rho$ are
indistinguishable as far as the
expectation value of $A$ is concerned,
yet the two states are clearly
distinguishable (with high 
statistical significance)
by the frequencies of observing the different
measurement outcomes $a_\nu$ when repeating the same 
measurement sufficiently often \cite{sho11}.

Here and in the following, the
term ``repetition'' (of a measurement) 
has the usual meaning, namely to perform 
a measurement of the same observable 
on an ensemble of systems in the same 
quantum mechanical state, each of them 
resulting in a random measurement outcome
and a concomitant collapse of the 
system state according to the common
rules of quantum mechanics,
see also above Eq. (\ref{5}).
%If so, it would no longer be right to say
%that the system equilibrates ``in practice'', 
%i.e., apart from unmeasurably small 
%deviations or extremely rare times.

Put differently, the actual problem 
of equilibration is to show that the 
frequencies, with which
the different possible measurement
outcomes are realized in the true
system state $\rho(t)$, are not
incompatible in any statistically 
significant way with the 
approximation $\bar\rho$.

In this statement of the problem, 
%the following two assumptions 
%are tacitly taken for granted:
%
%(i) The time point $t$ is considered 
$t$ is tacitly considered
as being chosen arbitrarily 
but then kept fixed.
(In Sect. \ref{s5} we will extend
the scope of our results also to cases 
when the measurement is taken at a 
different time in each repetition).
Moreover, the
%(ii) The 
number of repetitions,
%of the experiment, 
henceforth denoted as
$N_{rep}$, must remain reasonable, 
say
\begin{eqnarray}
N_{rep}<10^{30} \ .
\label{16}
\end{eqnarray}
(This bound is reached for $10^{12}$ 
repetitions per second during the age 
of the universe.)
To understand why such an upper 
bound is needed, we focus on the
generic case that
the probabilities (\ref{5}) and
(\ref{15})
with which the outcomes $a_\nu$ are
realized in the two states $\rho(t)$ and
$\bar\rho$, respectively,
%Disregarding the trivial
%and non-generic case 
%that these probabilities
%happen to exactly coincide 
are not exactly identical
for all $\nu$.
In the limit $N_{rep}\to\infty$
it then must become apparent with 
arbitrary statistical significance
that the approximation $\bar\rho$ 
is incompatible with the observed 
measurement outcomes,
which are sampled according to 
the true system state $\rho(t)$.
In other words, without imposing 
any upper bound on $N_{rep}$, 
the two states $\rho(t)$ and
$\bar\rho$ would generically 
be trivially distinguishable.

Important first steps in resolving the
above stated problem of equilibration
have been achieved in Ref. \cite{sho11}.
In doing so, the distinguishability of 
$\rho(t)$ and $\bar\rho$ was quantified 
as follows: 
Imagine that one of the two states 
$\rho(t)$ and $\bar\rho$ were randomly 
chosen with probability $1/2$ and then
used to sample one of the different
outcomes $a_\nu$ according to the
corresponding probabilities in 
(\ref{5}) or (\ref{15}). 
Now, the task is to guess from the
observed $a_\nu$ which state 
($\rho(t)$ or $\bar\rho$) has been 
chosen, and the probability that
this guess is correct was 
shown in Ref. \cite{sho11}
to be bounded by
%solely from the information which 
%outcome $a_\nu$ has been realized.
%Finally, it is shown in \cite{sho11} that 
%-- no matter what ``strategy'' 
%is adopted to guess the correct state --
%the  resulting probability
%the maximum success probability is 
$1/2+\sum_{\nu=1}^{N_A}|\kk_\nu(t) - \bar \kk_\nu|/4$.
Hence, the latter quantity was adopted in 
Ref. \cite{sho11} as the basic measure 
to quantify the distinguishability of $\rho(t)$ and 
$\bar\rho$ by means of $A$.
%in Ref. \cite{sho11}.
%as the distinguishability of 
%$\rho$ and $\bar\rho$.

We think that this approach is still 
unsatisfying in two respects:
(i) The underlying ``state guessing task''
is not exactly equivalent to the
above formulated ``actual problem 
of equilibration'':
In the actual problem, the outcomes are
always generated by $\rho(t)$, and not by
either $\rho(t)$ or $\bar\rho$ with equal 
probability.
Moreover, the actual task is not
to guess which of the two states was 
realized but rather to quantify the 
compatibility of the state $\bar\rho$ 
with the observed measurement 
outcomes.
(ii) The entire approach is limited to single shot
measurements, i.e. to $N_{rep}=1$. 
Indeed, already in the case of two 
repetitions of the same measurement, 
resulting in two outcomes $a_{\nu_1}$ 
and $a_{\nu_2}$, it is not clear at all
which of the two states should be guessed
according to the above described 
strategy from Ref. \cite{sho11}.
While a restriction like in (\ref{16})
still covers all cases of practical 
relevance, the same is no longer 
true for the restriction 
$N_{rep}=1$ considered in \cite{sho11}.

The main objective of our present paper
is to resolve the above issues (i) and (ii).
%is to overcome the above two shortcomings
%(i) and (ii) of the approach from \cite{sho11}.
In doing so, we will even admit one 
more generalization.
Namely, in every repetition $j$,
a measurement of a different
observable $A_j$ may be performed
(but the system state remains
the same in each repetition).
In particular, some or even all  
$A_j$ may still be identical,
but in general they are 
admitted to be different.
In doing so we denote --
similarly as in (\ref{5}) -- 
by $\KK_\nu^{(j)}$ 
the projectors onto the
eigenspaces of $A_j$ and by $a_\nu^{(j)}$
the corresponding eigenvalues,
where $\nu=1,...,N_{A_j}$.
Furthermore, their probabilities
of occurrence are denoted --
similarly as in (\ref{5}) 
and (\ref{15}) -- as
\begin{eqnarray}
\kk^{(j)}_\nu(t) :=  \tr\{\rho(t) \KK^{(j)}_\nu\}
\label{17a}
\end{eqnarray}
if the system is in the state $\rho(t)$ and as
\begin{eqnarray}
\bar \kk^{(j)}_\nu  :=   \tr\{\bar\rho\, \KK^{(j)}_\nu\}
\label{17b}
\end{eqnarray}
with respect to the state $\bar\rho$.
%in the hypothetical case that
%the system were in the state $\bar\rho$.
Accordingly, the outcome of our $N_{rep}$ 
measurements can be uniquely specified by 
a $N_{rep}$-dimensional
vector $\vec s$, whose $j$-th component 
$s_j \in \{1,...,N_{A_j}\}$ specifies which
outcome of $A_j$ was realized in the $j$-th 
measurement.
The probability to obtain the outcome 
$\vec s$ then follows as
\begin{eqnarray}
p_t(\vec s):=\prod_{j=1}^{N_{rep}} \kk^{(j)}_{s_j}(t)
\label{17}
\end{eqnarray}
if the system is in the state $\rho(t)$, and as
\begin{eqnarray}
\bar p(\vec s):=\prod_{j=1}^{N_{rep}} \bar \kk^{(j)}_{s_j}
\label{18}
\end{eqnarray}
with respect to the state $\bar \rho$.
Finally, with the definition
\begin{eqnarray}
N_{obs}:=\max_j N_{A_j}
\label{19}
\end{eqnarray}
we can conclude from (\ref{6}) that
\begin{eqnarray}
N_{A_j} \leq N_{obs} < 10^{20}
\label{20}
\end{eqnarray}
for all $j=1,...,N_{rep}$.

%%%%%%%%%%%%%%%%%%%%%%%%%%%%%%%%%%%%%%%%%%%%%%%%
\section{Distinguishability measures}
\label{s4}
In essence, the situation encountered 
above is as follows:
There is a true state $\rho(t)$
($t$ arbitrary but fixed)
and an approximative
state $\bar\rho$.
A series of $N_{rep}$ measurements
of the observables $A_j$ ($j=1,...,N_{rep}$)
is performed on the system state $\rho(t)$, 
resulting in one of the possible 
outcomes $\vec s$.
Within this setup, our key 
question is:
Given the outcome $\vec s$ of such a 
measurement series, 
does the approximation $\bar\rho$ 
explain the observed data 
$\vec s$ notably worse than the 
truth $\rho(t)$ would explain them, 
or are $\rho(t)$ and $\bar\rho$ 
both about equally well 
(or badly) compatible
with the given data $\vec s\,$?
If the latter is the case with
very high probability when
the entire measurement series
is repeated many times (i.e. each outcome
$\vec s$ is realized with 
probability $p_t(\vec s)$ from 
(\ref{17})), then the two 
states $\rho(t)$ and $\bar\rho$ 
are practically indistinguishable
by means of the considered 
observables $A_j$.
Put differently, the approximation $\bar\rho$ 
is as good as it possibly can 
be since it explains the observed
measurement outcomes 
practically as well as the 
best possible theory $\rho(t)$
would explain them.

To further substantiate these ideas, 
let us focus on an arbitrary but 
fixed ``test'' (or 
%``decision'',  
``rule'', ``strategy'', 
``criterion'', etc.)
by means of which we can (or hope to)
quantify (in whatever way)
how much worse (or possibly better) 
the approximation $\bar\rho$ is compatible
with a given data set $\vec s$ 
than $\rho(t)$.
In doing so, the two 
states $\rho(t)$ and $\bar\rho$ 
are thus considered as known.
In particular, all the 
probabilities from (\ref{17}) 
and (\ref{18}) are explicitly 
available and may be exploited 
by our test at hand.
In other words, $\rho(t)$, $\bar\rho$, and 
$\vec s$ are the input of the test,
which then acts like a black box
to produce an output in the form
of a real number 
$\D (\rho(t),\bar\rho,\vec s)$.
%(discriminator-function).

Without any significant loss of generality 
we assume that $\D (\rho(t),\bar\rho,\vec s)$ 
is standardized so that it only takes 
values within the interval $[-1,1]$.
Furthermore, $\D (\rho(t),\bar\rho,\vec s )=0$ 
indicates that $\rho(t)$ and $\bar\rho$ are 
(approximately) 
equally well (or badly) 
compatible with $\vec s$.
Finally, increasingly positive $\D$-values 
correspond to an increasing superiority
of $\rho(t)$ over $\bar\rho $ in explaining
the data $\vec s $, 
and likewise for negative $\D$-values.

Note that we can never be ``100\% sure'' 
that $\bar\rho$ is incompatible with 
the data $\vec s$
\footnote{At first sight, one might
think that $\bar p(\vec s )=0$
implies ``for sure'' that $\bar\rho$ 
is incompatible with the data $\vec s$.
However, as shown in Appendix A
(see below Eq. (\ref{y2})), 
$\bar p(\vec s )=0$ implies
$p_t (\vec s )=0$, i.e. such an
$\vec s$ is never realized.}.
Hence, any (reasonable) test
can only make certain probabilistic 
statements (based on some certain notion 
of probability, likelihood, 
confidence, plausibility, ...) 
about the compatibility of $\bar\rho $ 
with $\vec s $, and likewise for $\rho(t)$.

We also note that for some ``strange'' 
(unlikely but not impossible) 
outcomes $\vec s$ of the measurement series, 
even the ``reality'' $\rho(t)$
may be incompatible 
(in the above mentioned probabilistic sense)
with $\vec s$ according 
to the criteria of the given test.
Likewise, the compatibility
of certain $\vec s$ with $\bar\rho $ 
may actually be better (or less bad)
than with $\rho(t)$.
Intuitively (or from a Bayesian viewpoint), 
it seems plausible that such cases may 
be realized whenever a given outcome $\vec s$
has the property that
$\bar p(\vec s)>p_t (\vec s)$.
Quantitatively such cases are taken into account
by the negative $\D$ values.

A particularly simple example is
\begin{eqnarray}
\!\!\!\!\!\!\!\!\!\!\!\!
\D (\rho(t),\bar\rho ,\vec s) &  := & f\!\left[
p(\rho(t) | \vec s) - p(\bar\rho | \vec s)\right]
\label{21}
\\
\!\!\!\!\!\!\!\!\!\!\!\!
p(\rho(t) | \vec s)  & := & \frac{p_t(\vec s)}{p_t(\vec s)+\bar p(\vec s)}
\label{22a}
\\
p(\bar\rho | \vec s)  & := & \frac{\bar p(\vec s)}{p_t(\vec s)+\bar p(\vec s)} \ ,
\label{22b}
\end{eqnarray}
where $f[x]$ is some monotonically increasing 
functions of $x$ with $f[-1]=-1$, $f[0]=0$,
$f[1]=1$, for instance $f[x]=x$.
(We tacitly restrict ourselves to outcomes 
$\vec s $ which are realized with non-vanishing
probability, hence the denominators in 
(\ref{22a}) and (\ref{22b}) are non-zero.
We also note that since 
$p(\bar \rho | \vec s)=1-p(\rho (t)| \vec s)$,
the right hand side of (\ref{21}) could equally well 
be written as a function of $p(\rho (t)| \vec s)$ alone.)
Within the framework of Bayesian inference, 
%and choosing $f(x)=g(x)=x$,
$p(\rho(t) | \vec s)$ and $p(\bar\rho | \vec s)$ in (\ref{22a}) 
and (\ref{22b})
can be identified with the posterior probabilities of 
$\rho(t)$ and $\bar\rho$, given $\vec s $ has been observed,
%normalized so that the total
%probability $\kk_1(\vec s ) + \kk_2(\vec s )$ 
%equals unity,
and assuming equal prior probabilities
for $\rho(t)$ and $\bar\rho$
(i.e., before the observations $\vec s $ are available).
Accordingly, (\ref{21}) may be interpreted
as quantifying the likelihood of
$\rho(t)$ compared to that of $\bar\rho $.
We, however, remark that even without adopting a 
Bayesian viewpoint, all quantities 
in (\ref{21})-(\ref{22b}) remain well 
defined and admit a decent probabilistic 
interpretation.

Analogously as in (\ref{21}), one also could, for example, 
quantify the compatibility of $\rho(t)$ and $\bar\rho$ 
with $\vec s$ in cases when all $A_j$ are equal
by defining $\D (\rho(t),\bar\rho ,\vec s)$ in some
suitable way via the two values which are obtained 
by applying a $\chi^2$ test to the two 
``null hypotheses'' 
%deriving from 
$\rho(t)$ and $\bar\rho$.

Returning to the general case of an 
arbitrary but fixed test,
any given such test may still
admit many different reasonable
choices of $\D$ (e.g. different 
functions $f$ in (\ref{21})).
Our first key hypothesis is now that for any
given test it is possible to choose
a function $\D$
so that the quality of this 
test is reasonably quantified by
the distinguishability measure
\begin{eqnarray}
\QQ (t):=\sum\limits_{\vec s}  p_t(\vec s ) \, \D (\rho(t),\bar\rho ,\vec s ) 
\ ,
\label{23}
\end{eqnarray}
i.e., by averaging $\D (\rho(t),\bar\rho ,\vec s )$ over 
many measurement series and weighting 
every possible outcome $\vec s$ with the frequency
$p_t(\vec s)$ with which it is realized.
More precisely, the existence of at least one
$\D $-function is postulated for which a $\QQ $-value 
close to unity indicates that the given test 
quite reliably recognizes the incompatibility 
of $\bar\rho $ with the measurement series $\vec s$ 
(which was sampled according to $\rho(t)$), 
while a $\QQ $-value close to zero
indicates that there is no way to
recognize by means of the given test 
any significant difference between
the truth $\rho(t)$ and the approximation 
$\bar\rho $ when sampling a data set 
$\vec s $ according to $\rho(t)$.
Note that small negative $\QQ $ still 
indicate a good compatibility
of $\bar\rho$ with the measurements, 
while non-small negative 
$\QQ $-values are also possible but would 
quite plausibly indicate that the given 
test is futile, or that one rather should 
employ $-\D $ instead of $\D $.

%Note that $\D (\rho(t),\bar\rho ,\vec s )\in [-1,1]$
%implies $\QQ \in[-1,1]$ and that we can restrict 
%ourselves without loss of generality to 
%cases with 
%\begin{eqnarray}
%\QQ  \geq 0 \ .
%\label{2}
%\end{eqnarray}
%The reason is that in case of $\QQ <0$ we can
%consider $-\D $ instead of $\D $ as our 
%pertinent test quantifier.

The complete set of all tests which 
fulfill our above assumptions 
may still admit some undesirably 
biased $\D $-functions.
The most trivial example is
$\D (\rho(t),\bar\rho ,\vec s ):=1$ for 
all $\vec s $, yielding the highest 
possible score of $\QQ (t)=1$.
The only viable way out seems 
to admit only tests, whose $\D $-function
does not exploit the information 
that the data $\vec s$ were actually
sampled according to $\rho(t)$.
In particular, we may imagine 
(as a hypothetical Gedankenexperiment) 
that the system was not in the state 
$\rho(t)$ but rather in the state 
$\bar\rho$ without telling this fact
to the person working with a given 
$\D$-function.
If we would in this way
secretly sample $\vec s$ not
according to $\rho(t)$ but 
rather according to
$\bar\rho$, then the very same 
$\D $-function
should now be able to recognize 
that the data are (on the average)
better explained by $\bar\rho$ 
than by $\rho(t)$.

Therefore, our second key hypothesis
is that it is sufficient to focus on 
tests, whose $\D $-functions satisfy
the following additional symmetry property:
Imagine many repetitions of 
our so far considered measurement series.
But now, in every repetition, 
either $\rho(t)$ or $\bar\rho$
is randomly selected with
probability $1/2$ and then used to 
randomly generate 
(via the usual quantum mechanical
measurement process) a measurement 
series $\vec s$ according to the
corresponding probability (\ref{17}) 
or (\ref{18}).
%However, in doing so, the 
%$\D$-function does not know which state 
%has been selected in any given realization.
In this case, we require that the $\D$-function 
must be unbiased on the average, i.e.
\begin{eqnarray}
\sum_{\vec s} \frac{p_t(\vec s )+\bar p(\vec s )}{2}\, 
\D (\rho(t),\bar\rho ,\vec s )=0 \ .
\label{24}
\end{eqnarray}

While the above two hypotheses seem 
difficult or impossible to derive from 
some more fundamental principles, 
they appear quite reasonable as they 
stand and are thus taken for granted 
from now on.
Their most important virtue is that
they imply -- as demonstrated in detail
%As a consequence, we derive
in Appendix B -- the general 
rigorous bound
\begin{eqnarray}
|\QQ (t)| \leq 
\QQ _{max}(t):=
\frac{1}{2} \sum\limits_{\vec s}
|p_t(\vec s ) - \bar p(\vec s )|  
%=:\QQ _{max}(t)
\ .
\label{25}
\end{eqnarray}
for the distinguishability 
measure $\QQ (t)$ from (\ref{23}),
independently of any further
peculiarities of the considered test
and the concomitant 
$\D $-function.
%for any pertinent test which satisfies 
%the two above discussed hypotheses.
This is the first main result of our paper.

In particular, one readily verifies that
the example from (\ref{21})-(\ref{22b})
with $f[x]=x$
exhibits the symmetry (\ref{24}) 
and respects the bound (\ref{25}).
More generally, the symmetry (\ref{24})
imposes a non-trivial constraint
on $f[x]$ and 
a direct verification of (\ref{25})
(without recourse to Appendix B)
becomes difficult.

In hindsight, the original task to quantify 
the compatibility of $\bar\rho$ with the
measurement series $\vec s$ 
may have appeared quite daunting since
this can be done in so many different 
ways, most of which one possibly
did not even think of in the first place.
The appeal of our main result (\ref{25}) 
is that it applies independently of
the concrete manner in which
the compatibility is quantified.
In particular, no explicit knowledge 
is needed about the best possible 
way to quantify the compatibility
of $\bar\rho$ with $\vec s$.
The only, very weak and plausible
requirements are that all considered
distinguishability measures can be written 
in the form (\ref{23}) for some suitable $\D$ 
function, and that they respect the 
symmetry condition (\ref{24}).

%%%%%%%%%%%%%%%%%%%%%%%%%%%%%%%%%%%%%%%%%%%%%%%%
\section{Final Result and Conclusions}
\label{s5}
The upshot of the previous 
section is: 
If we can show that $\QQ _{max}(t)$ from (\ref{25}) 
is a small quantity, then there is no way to 
experimentally detect any statistically 
significant deviation of the approximation 
$\bar\rho$ from the true system state 
$\rho(t)$.
The latter statement applies
for an arbitrary but fixed time $t$
and for an arbitrary but
fixed measurement series
 $A_1,...,A_{N_{rep}}$.
Hence, if we can show that the 
same statement holds simultaneously
for all measurement series which
satisfy (\ref{16}) and (\ref{20})
and for the overwhelming majority
of all sufficiently late times $t$ 
then it follows 
-- analogously as in Sect. \ref{s3a} --
that approximating $\rho(t)$ by 
$\bar\rho$ can be considered as 
perfect for all practical 
purposes \footnote{We recall that (\ref{20})
follows from our assumption (\ref{6})
and the definition (\ref{19}).}.

As detailed in Appendix C, it is indeed
possible to show that a result of the
above type holds true.
Quantitatively, the result is analogous 
to Eqs. (\ref{13}), (\ref{14}), stating that
\begin{eqnarray}
T^\ast/T & \leq & \epsilon
\label{26}
\\
T^\ast & := & 
\big| 
\{\, 0 \leq t  \leq T\, : \, \QQ _{max}(t) > \epsilon\, \}
\big| 
\label{28}
\\
\epsilon & := & (122\, g\, {\max_n}' p_n)^{1/4}\, N_{obs}^{1/2}\, N_{rep}
\label{27}
\end{eqnarray}
for all sufficiently large $T$.
This is the main final 
result of our paper.
Its discussion 
can be conducted along very similar lines as in
Sect. \ref{s3}, hence we only recapitulate here
the main points:
On the right hand side of (\ref{27}),
$g$ denotes the maximal degeneracy 
of energy gaps from (\ref{12})
(with $g=1$ for Hamiltonians with
a generic spectrum).
Furthermore, $\slp$ is the second 
largest level population and, according to
 (\ref{8}), is typically exponentially small in $f$
for a system with $f\gg 1$ degrees of freedom. 
In view of (\ref{16}) and (\ref{20}) we thus can
conclude that $\epsilon$ in (\ref{27}) becomes 
an extremely small number already for 
systems with, say, 
more than $10^3$ degrees of freedom.
In turn, the Lebesgue measure (\ref{28}) of those
times $t \in [0,T]$, for which there possibly 
may exist a non-negligible chance to observe
a resolvable difference between $\rho(t)$ 
and $\bar\rho$ by some suitable 
measurement procedure, is -- according to (\ref{26})
-- negligibly small compared to all times
$t\in[0,T]$, provided $T$ is sufficiently large.

To summarize, the steady state $\bar\rho$
approximates the true
state $\rho(t)$ practically perfectly 
for all sufficiently large times $t$.
While the two states are rigorously 
speaking never close to each other 
in some mathematically obvious way, 
the observable differences are either 
unresolvably small or negligibly
rare from all practical 
%(experimental) 
points of view.

We finally note that by admitting the possibility 
to employ in every repetition of the experiment
a different observable $A_j$,
our approach actually also covers
the case when the measurement is 
performed in every repetition at a 
different time point
(which strictly speaking applies to 
every real experiment).
The reason is the usual equivalence
of the Schr\"odinger and Heisenberg
pictures, i.e., a temporal change of the 
system state can be replaced by an
equivalent change of the considered 
observable.

%%%%%%%%%%%%%%%%%%%%%%%%%%%%%%%%%%%%%%%%%%%%%%%%
\subsection*{Acknowledgments}
Special thanks are due to Michael Kastner 
for invaluable discussions and 
very insightful suggestions.
This work was supported by 
DFG-grant RE1344/7-1.

%%%%%%%%%%%%%%%%%%%%%%%%%%%%%%%%%%%%%%%%%%%%%%%%
\section*{Appendix A}
\label{a1}
In this Appendix, we provide the derivation
of Eqs. (\ref{10},\ref{11}),
i.e. we show that for all sufficiently large $T$
\begin{eqnarray}
\int_0^T\!  \frac{dt}{T}\, [\sigma(t)]^2 
& \leq &
%24 
3 \, \tr\{\bar\rho A^2\}  \, g\, {\max_n}' p_n
%6 \, \da^2 \, g\, {\max_n}' p_n
\label{1a}
\\
\sigma(t) & := & \tr\{\rho(t)  A\} - \tr\{\bar\rho A\} \ ,
\label{1a1}
%\left[\langle A\rangle_{\!\rho(t)} - \langle A\rangle_{\!\bar\rho}\right]
\end{eqnarray}
where 
%$\da$ is the difference between the
%largest and smallest eigenvalues of $A$,
$g$ is the maximal degeneracy of energy 
gaps from (\ref{12}), and $\slp$
is the second largest among all 
energy level populations from (\ref{7}).
%To a large extent, the calculations
%in this Appendix amount to a  
%straightforward extension and unification
%of Refs. \cite{sho12,rei12,het15}.
%Yet, in order to make the paper 
%self-contained,
%we are providing the entire streamlined 
%and partially complemented line of 
%reasoning.

%%%%%%%%%%%%%%%%%%%%%%%%%%%%%%%%%%%%%%%%%%%%%%%%
\subsubsection*{Preliminaries} 
\label{a10}
We recall that the $P_n$ in (\ref{1}) are the 
projectors onto the eigenspaces of the 
Hamiltonian $H$, where $n$ runs from
1 to infinity or to some upper finite limit.
In other words, $n\in I$, where the index
set $I$ is either equal to $\NN$ or
of the form $\{1,...,L\}$ with a finite 
upper limit $L\in\NN$.
Hence, the projectors $P_n$ satisfy the usual
orthogonality and completeness relations, i.e.,
\begin{eqnarray}
P_m P_n & = & \delta_{mn} P_n 
\label{x1}
\end{eqnarray}
for all $m,\,n\in I$ and
\begin{eqnarray}
\sum_n P_n =1 \ ,
\label{x0}
\end{eqnarray}
where $1$ is the identity operator
and, as in (\ref{1}), the sum runs over
all $n\in I$.
Next we define
\begin{equation}
\rho_n:=P_n\rho(0)P_n
\label{x2}
\end{equation}
and we denote by $S_O$ the set of operators
which contains all the $P_n$'s and all the
$\rho_n$'s.
It follows that all operators in $S_O$ are
Hermitian and commute with each other, 
hence there must be a common eigenbasis 
for all operators contained in $S_O$.
In other words, there is an orthonormalized
basis $\{|\gamma\rangle\}_{\gamma=1}^D$,
whose dimension $D$ may be either finite 
or infinite, and whose associated projectors 
$P_\gamma:=|\gamma\rangle\langle\gamma|$
commute with all operators contained in $S_O$.
%As a consequence, all P_\gamma$ also commute 
%with $H$ from (\ref{1}), i.e. $|\gamma]\rangle$
%are eigenvectors of $H$.
As a consequence, also the projectors 
\begin{eqnarray}
X & := & \sum_{\gamma=1}^{d}|\gamma\rangle \langle\gamma| 
\,,
\label{:2A}
\\
Y & := & 1-X = \sum_{\gamma=d+1}^{D}|\gamma\rangle \langle\gamma| 
\, .
\label{:3A}
\end{eqnarray}
commute with all $P_n$ and $\rho_n$, i.e.
\begin{equation}
XP_n=P_nX
\, , \   
X \rho_n=\rho_n X
\label{x3}
\end{equation}
for all $n$, and likewise for $Y$.
With the definition
\begin{equation}
\tilde A  :=  XAX
\ .
\label{:10A}
\end{equation}
for arbitrary observables $A$, 
it follows with (\ref{x3}) that
\begin{equation}
\tilde{P}_{n} :=  XP_{n}X = P_{n}X = XP_{n} \ .
\label{z5}
\end{equation}
In particular,
if one defines $\tilde A_{mn}$
analogously as in (\ref{3}) 
%as $\tr\{P_m\rho(0)P_n \tilde A\}$,
then one readily verifies that
\begin{eqnarray}
\tilde A_{mn} & := & \tr\{P_m\rho(0)P_n \tilde A\}=\tr\{ \tilde P_m \rho(0) \tilde P_n A \}
\label{a40} 
%\\
%\tilde\rho_{mn} & := & \tilde P_m \rho(0) \tilde P_n
%\ .
%\label{a41}
\end{eqnarray}
by exploiting (\ref{:10A}), (\ref{z5}), and the cyclic 
invariance of the trace.

For any given basis vector $|\gamma\rangle$, 
one can infer from (\ref{x1}) and (\ref{x0})
that $P_n|\gamma\rangle$ equals 
$|\gamma\rangle$ (eigenvalue $1$)
for exactly one index $n$, and 
equals the null vector $|0\rangle$ 
(eigenvalue $0$)
for all other indices $n$.
Since $d$ in (\ref{:2A}) is finite,
it follows that at least one and 
at most $d$ among all the
$\tilde P_n$'s in (\ref{z5})
are {\em not} identically zero.
Without loss of generality, we can and
will choose the labels $n$ and some 
suitable integer $N\in\{1,...,d\}$ 
so that $\tilde{P}_{n}$ 
is non-zero if and only if 
$n\in\{1,...N\}$.
Furthermore, we can and will
choose the labels $\gamma$ 
and $n$ so that the energy level
populations $p_n$ from (\ref{7}) assume
their maximal value for $n=1$, i.e.,
\begin{equation}
\max_{n} p_n = p_1 \ .
\label{x3a}
\end{equation}
Accordingly, the second largest level
population can be written as
\begin{equation}
{\max_{n}}' p_n = \max_{n\geq 2} p_n \ .
\label{x3b}
\end{equation}

It follows that 
%$\tilde\rho_{mn}$ and thus 
$\tilde A_{mn}$ in (\ref{a40})
must vanish unless $m \leq N$ and 
$n\leq N$.
With (\ref{2}) we thus can conclude that
\begin{eqnarray}
\tr\{\rho(t)\tilde A\} = \sum_{m,n=1}^N
\tilde A_{mn} \, e^{i (E_n-E_m) t} \ .
\label{z2}
\end{eqnarray}
Likewise,  one can infer from (\ref{9}) and 
(\ref{:10A}) that
\begin{eqnarray}
\tr\{\bar\rho \tilde A\} = \sum_{n=1}^N \tilde A_{nn}  \ .
\label{z2a}
\end{eqnarray}
Together, we thus obtain
\begin{eqnarray}
\tilde\sigma(t)  & := & \tr\{\rho(t)\tilde A\} - \tr\{\bar\rho \tilde A\} 
\nonumber
\\
& = & 
\sum_{m\not =n}^N
\tilde A_{mn} \, e^{i (E_n-E_m) t}\ ,
\label{z3}
\end{eqnarray}
where the sum runs over all $m,n\in\{1,...,N\}$ with $m\not=n$.

Since $\rho(0)$ is a Hermitian, non-negative operator, 
there exists a Hermitian, non-negative operator, 
which we denote by $\rho^{1/2}$, and which satisfies 
the relation $\rho^{1/2}\rho^{1/2}=\rho(0)$. 
With the Cauchy-Schwarz inequality 
\begin{equation}
\left|\tr\{B^{\dagger}C\}\right|^{2}
\leq
\tr\{B^{\dagger}B\}\tr\{C^{\dagger}C\}
\label{:11A}
\end{equation}
for the scalar product $\tr\{B^{\dagger}C\}$ of 
arbitrary operators $B$ and $C$ 
(for which all traces in (\ref{:11A}) exist), 
and exploiting $P_m=P_m^2$ (cf. (\ref{x1}))
and the cyclic invariance of the trace,
we can conclude from (\ref{a40}) that
\begin{eqnarray}
|\tilde A_{mn}|^2 & = & \tr\{(P_m \rho^{1/2})(\rho^{1/2} P_n \tilde A P_m) \}
\nonumber
\\
& \leq &
\tr\{ P_m \rho(0) P_m\}\, \tr\{ P_n \rho(0) P_n \tilde A P_m \tilde A \} \ .
\label{x4}
\end{eqnarray}
The first factor in the last line can be identified 
with the level population $p_m$ from (\ref{7}). 
In combination with (\ref{x2}), (\ref{:10A}), and (\ref{z5}) 
we thus obtain
\begin{eqnarray}
|\tilde A_{mn}|^2 
& \leq & 
p_m\, \tr\{ \rho_n X A X P_m X A X\}
\nonumber
\\
& = &
p_m\, \tr\{ X \rho_n A \tilde P_m A\} \ .
\label{x5}
\end{eqnarray}

Next, we observe that $A \tilde P_m A$ as well as
$\rho_n$ from (\ref{x2}) are both Hermitian, non-negative  
operators, and that every $|\gamma\rangle$
is an eigenvector of both $\rho_n$ and $X$
from (\ref{:2A}).
Upon employing the basis $|\gamma\rangle$ 
to evaluate the trace in (\ref{x5}), one thus 
obtains
\begin{eqnarray}
|\tilde A_{mn}|^2 
& \leq & 
p_m\, \tr\{\rho_n A \tilde P_m A\} \ .
%p_m\, \tr\{\rho_n \tilde A P_m \tilde A\} \ .
\label{z8}
\end{eqnarray}
With $P_m^2=P_m$ (cf. (\ref{x1})) and (\ref{z5})
we can infer that $\tilde P_m=P_mXP_m$.
Exploiting the cyclic invariance of the trace we 
thus can rewrite the last factor in (\ref{z8}) as
$\tr\{X B \}$ with $B:=P_m A \rho_n A P_m$.
Since $B$ is a Hermitian, non-negative 
operator, it follows that
$\tr\{X B \}\leq \tr\{B \}$
and by the same steps as before that
$\tr\{B \}=\tr\{\rho_n A P_m A\}$.
Altogether, we finally obtain
\begin{eqnarray}
|\tilde A_{mn}|^2 
& \leq & 
p_m\, \tr\{\rho_n A P_m A \} \ .
%p_m\, \tr\{\rho_n \tilde A P_m \tilde A\} \ .
\label{z8a}
\end{eqnarray}

%%%%%%%%%%%%%%%%%%%%%%%%%%%%%%%%%%%%%%%%%%%%%%%%
\subsubsection*{Step 1} 
\label{a11}
In this subsection,
we closely follow the line of 
reasoning from Ref. \cite{rei12}.
The main result will 
be (\ref{a23}).

Due to (\ref{:2A}) and (\ref{:3A}) it follows 
for arbitrary density operators $\rho$ and 
observables $A$ that 
\begin{align}
\tr\{\rho A\} 
& =\tr\{\left(X+Y\right)\rho\left(X+Y\right)A\}\nonumber \\
& =R_{1}+R_{2}+R_{3}
\label{:4A}
\\
R_{1} &  := \tr\{X\rho XA\}\,,
\label{:5A}
\\
R_{2} &  := \tr\{Y\rho \left(X+Y\right) A\} = \tr\{Y\rho A\}\,,
\label{:6A}
\\
R_{3} &  := \tr\{X\rho YA\}\,.
\label{:7A}
\end{align}
Exploiting the cyclic invariance of the trace 
and the definition (\ref{:10A}) yields
\begin{align}
R_{1} & =\tr\{\rho\tilde{A}\}
\ .
\label{:8A}
\end{align}

By a similar line of reasoning as in the derivation of
(\ref{x4}) we can rewrite (\ref{:6A}) as
\begin{align}
|R_{2}|^{2} 
%& =|\tr\{(Y\rho^{1/2}) (\rho^{1/2}A)\}|^{2}
%\nonumber 
%\\
& \leq\tr\{Y\rho Y\}\, \tr\{A\rho A\} 
\ .
\label{:12A}
\end{align}
Exploiting that for arbitrary Hermitian, 
non-negative operators $B$ and $C$ 
\begin{equation}
\tr\{BC\}\leq\left\Vert B\right\Vert \tr\{C\} \ ,
\label{:13A}
\end{equation}
where $\left\Vert B\right\Vert$ denotes the
standard operator norm (largest eigenvalue),
the last term in (\ref{:12A}) can be rewritten as
\begin{equation}
\tr\{A\rho A\}=\tr\{\rho A^{2}\}
\leq
\left\Vert A\right\Vert ^{2}\tr\{\rho\}=\left\Vert A\right\Vert ^{2} <\infty
\ .
\label{:15A}
\end{equation}
The last inequality follows from (\ref{4}) and (\ref{6}).
Finally we can conclude from (\ref{:3A}) that 
\begin{equation}
\tr\{Y\rho Y\} = \tr\{Y\rho\}
=
\sum_{\gamma=d+1}^{D} \langle \gamma | \rho|\gamma\rangle 
\label{:16A}
\end{equation}
and hence from (\ref{:12A}) that
\begin{equation}
\left|R_{2}\right|^{2}\leq\left\Vert A\right\Vert ^{2}
\sum_{\gamma=d+1}^{D}\langle\gamma|\rho|\gamma\rangle \, .
\label{:17A}
\end{equation}
Analogously, one finds for $R_{3}$ 
from (\ref{:7A}) that
\begin{align}
\left|R_{3}\right|^{2} 
& \leq\tr\{Y\rho Y\}\left\Vert A\right\Vert^{2}
\tr\{X \rho X\}
\nonumber 
\\
& \leq\left\Vert A\right\Vert^{2}
\sum_{\gamma=d+1}^{D}
\langle\gamma|\rho|\gamma\rangle \,.
\label{:18A}
\end{align}
Introducing (\ref{:8A}), (\ref{:17A}), and (\ref{:18A}) 
into (\ref{:4A}) finally yields
\begin{equation}
\left|\tr\{\rho A\}-\tr\{\rho\tilde{A}\}\right|
\leq
2\left\Vert A\right\Vert
\left(\sum_{\gamma=d+1}^{D}
\langle\gamma|\rho|\gamma\rangle \right)^{1/2}
\label{:19A}
\end{equation}
for arbitrary density operators 
$\rho$. 

Next we focus on the two specific
density operators $\rho(t)$ 
and $\bar{\rho}$ from sections 
\ref{s2} and \ref{s3a}.
%and define their counterparts
%$\tilde\rho(t)$ and $\tilde{\bar\rho}$
%analogously as in (\ref{:9A}).
With the definition
\begin{eqnarray}
%\sigma (t) & := & \tr\{\rho (t)A\} - \tr\{\bar\rho A \}\,,
%\label{:19A-0}
%\\
%\tilde\sigma (t) & := &
%\tr\{\rho (t) \tilde A\}
%- \tr\{\bar\rho \tilde A \}\,,
%\label{:19A-1}
%\\
\delta (t) & := & \sigma(t)-\tilde\sigma(t) \,
\label{:19A-2}
\end{eqnarray}
it readily follows from (\ref{1a1}), (\ref{z3}), 
(\ref{:19A}), and (\ref{:19A-2}) that
\begin{eqnarray}
%|\sigma(t)|  & \leq &
%|\tilde\sigma(t)| + |\delta (t)|
%\label{:20A}
%\\
|\delta (t)| 
& = & |\tr\{\rho(t)A\} - \tr\{\bar\rho A\} 
- \tr\{\rho(t) \tilde A\} + \tr\{\bar\rho \tilde A\}|
\nonumber \\
& \leq & |\tr\{\rho(t)A\} - \tr\{\rho(t) \tilde A\}| 
+ |\tr\{\bar\rho A\} - \tr\{\bar\rho \tilde A\}|
\nonumber \\
& \leq & 4\left\Vert A\right\Vert
\left(\sum_{\gamma=d+1}^{D}
\langle\gamma|\bar\rho|\gamma\rangle 
\right)^{1/2} \ .
\label{:22A}
\end{eqnarray}
In the last step exploited that 
$\langle\gamma|\rho(t)|\gamma\rangle=
\langle\gamma|\bar\rho|\gamma\rangle$
for all $t$ and $\gamma$, 
as can be verified by choosing
$A=|\gamma\rangle\langle\gamma|$
in (\ref{2}) and (\ref{9}).

Observing that $\Vert A \Vert$ is 
finite (cf. (\ref{:15A})) and that
$\sum_{\gamma=1}^{d}\langle\gamma|\bar \rho|\gamma\rangle$
is monotonically increasing with $d$ and 
bounded from above by $\tr\{\bar\rho\}=1$, 
it follows from (\ref{:22A})
that for any given $\epsilon>0$ there 
exists a finite $d$ with
\begin{equation}
\delta^2(t) \leq \epsilon 
\label{a23}
\end{equation}
for all $t$.
This is the main result of the 
present subsection.

%%%%%%%%%%%%%%%%%%%%%%%%%%%%%%%%%%%%%%%%%%%%%%%%
\subsubsection*{Step 2} 
\label{a12}
In this subsection, we closely follow 
the line of reasoning from Ref. \cite{het15},
which in turn amounts to a simplification
of the previous approach from 
Refs. \cite{sho12}.
The main results will be
(\ref{:31A-3}) and (\ref{:33A-1}).

Denoting the set of unequal label pairs as
\begin{equation}
\mathcal{G} := \{ (m,n)\, :\, m,n\in\{1,...,N\},\, m\neq n\} 
\label{:28A}
\end{equation}
and defining for any $\alpha=\left(m,n\right)\in\mathcal{G}$
\begin{equation}
G_{\alpha} :=  E_{m}-E_{n}\, ,\ 
v_{\alpha} := \tilde A_{mn} \ ,
\label{:29A}
\end{equation}
it readily follows with (\ref{z3}) that
\begin{equation}
\tilde \sigma^{2}(t)
=
\left|\sum_{\alpha\in\mathcal{G}}
v_\alpha \, e^{-iG_\alpha t} 
\right|^{2} = \underset{\alpha,\beta\in\mathcal{G}}{\sum}
v_{\alpha}^{*} v_{\beta} \, e^{i (G_{\alpha}-G_{\beta}) t}
\label{:30A-3}
\end{equation}
and hence that
\begin{eqnarray}
\tilde \sigma^2(t) & = & R(t)+S
\label{:30A}
\\
R(t) & := &
\underset{\underset{G_{\alpha}\neq G_{\beta}}{\alpha,\beta\in\mathcal{G}}}{\sum}
v_{\alpha}^{*} v_{\beta}
%\int_{0}^{T}\frac{dt}{T}\,
e^{i (G_{\alpha}-G_{\beta}) t}
\label{:30A-1}
\\
S & := & \underset{\underset{G_{\alpha}=G_{\beta}}{\alpha,\beta\in\mathcal{G}}}{\sum}
v_{\alpha}^{*} v_{\beta} \ .
\label{:30A-2}
\end{eqnarray}
Note that both $R(t)$ and $S$ are real numbers 
and that their sum must be non-negative.

Abbreviating the time average of an 
arbitrary function $f(t)$ as
\begin{equation}
\left\langle f(t)\right\rangle_{T}
:= \int_{0}^{T}\frac{dt}{T} \, f(t) \ ,
\label{:26A}
\end{equation}
one readily finds by integrating over the 
exponential in (\ref{:30A-1})  that
\begin{equation}
\left\langle R(t)\right\rangle_{T} \leq\frac{1}{T}
\underset{\underset{G_{\alpha}\neq G_{\beta}}{\alpha,\beta\in\mathcal{G}}}{\sum}
\left|v_{\alpha}^{*} v_{\beta} \right|
\frac{2}{\left|G_{\alpha}-G_{\beta}\right|}
\ .
\label{:31A-1}
\end{equation}
Since the number of summands is 
finite (cf. (\ref{:28A})),
we can conclude from (\ref{:30A})-(\ref{:31A-1})
that for any given $\epsilon>0$ there 
exists a finite $T_\epsilon$ with
\begin{equation}
\left\langle \tilde \sigma^{2}(t)\right\rangle_{T}
\leq \epsilon + S
\label{:31A-3}
\end{equation}
for all $T\geq T_\epsilon$.

Next, we consider subsets $\mathcal{G}_j$ of $\mathcal{G}$ from
(\ref{:28A}), defined via the property that all elements $\alpha=(m,n)$
which belong to the same subset $\mathcal{G}_j$ exhibit identical
energy gaps $G_\alpha:= E_m-E_n$ (cf. (\ref{:29A})),
while for any pair $\alpha\in \mathcal{G}_j$, $\beta\in \mathcal{G}_k$
with $j\not=k$ the corresponding energy gaps 
$G_\alpha$ and $G_\beta$ are different.
It follows that the number of subsets $\mathcal{G}_j$ is finite,
say $j=1,...,J$, that $\mathcal{G}$ is the disjoint union
of all those subsets $\mathcal{G}_j$, and that
each subset $\mathcal{G}_j$ contains a finite number
of elements, which we denote by $g_j$.
Recalling that $g$ from (\ref{12}) 
denotes the maximal number of degenerate 
energy gaps it follows that
\begin{eqnarray}
g_j\leq g
\label{a30}
\end{eqnarray}
for all $j$.
Furthermore, we can rewrite $S$ from (\ref{:30A-2}) as
\begin{equation}
S=\sum_{j=1}^J\sum_{\alpha,\beta\in \mathcal{G}_j} 
v_{\alpha}^{*} v_{\beta} \ .
\label{:31A}
\end{equation}
Next we define the scalar product 
$\langle B|C \rangle:=\sum_{k,l=1}^M B_{kl}^{*}C_{kl}$
for arbitrary $M\times M$ matrices $B$ and $C$.
For the special choice $B_{kl}:=x_k$ (independent of $l$)
and $C_{kl}:=x_l$ (independent of $k$) the Cauchy-Schwarz
inequality implies
\begin{align}
\left| \sum_{k,l}^{M}x_{k}^{*} x_{l}\right|^2 
&
\leq \sum_{k,l}^{M} |x_k|^2\sum_{k,l}^{M} |x_l|^2
=\left(M \sum_{k}^{M} |x_k|^2\right)^2
\label{:32A}
\end{align}
for arbitrary complex numbers $x_1,...,x_M$.
Observing that the last sum
in (\ref{:31A}) is exactly of this structure
with $M=g_j$, it follows with (\ref{a30}) that
\begin{displaymath}
S \leq
\sum_{j=1}^J g_j   \sum_{\alpha\in \mathcal{G}_j}   |v_\alpha|^2
\leq
g   \sum_{j=1}^J   \sum_{\alpha\in \mathcal{G}_j}   |v_\alpha|^2
=
g   \sum_{\alpha\in \mathcal{G}}   |v_\alpha|^2
\, .
%\label{:33A}
\end{displaymath}
Returning to our original notation via (\ref{:28A}) and
(\ref{:29A}), we finally obtain
\begin{align}
S & \leq 
g \sum_{m\not = n}^N |\tilde A_{mn}|^2
\ .
\label{:33A-1}
\end{align}
This relation together with (\ref{:31A-3}) 
is the main result of the present subsection.

%%%%%%%%%%%%%%%%%%%%%%%%%%%%%%%%%%%%%%%%%%%%%%%%
\subsubsection*{Step 3} 
\label{a13}
In this subsection,
we closely follow the line of 
reasoning from Ref. \cite{rei12}.
The main result will be
(\ref{z14}).

Denoting by $\Sigma_{1}$ the partial sum on the right hand side
of (\ref{:33A-1}) over all summands with $n=1$ implies with 
(\ref{z8a}) that
\begin{eqnarray}
\Sigma_{1} & := & \sum_{m=2}^N |\tilde A_{m1}|^2
\leq
%\sum_{m=2}^N p_m \tr\{ \rho_1 A P_m A\}
%\nonumber
%\\
%& \leq &
\underset{n\geq 2}{\textrm{max}}\,p_{n} \, W_1
\label{z11a}
\\
W_1 & := & 
\sum_{m=2}^N \tr\{ \rho_1A P_m A\}
\ .
\label{z11b}
\end{eqnarray}
Since $\rho_nA P_m A$ is a non-negative 
operator for arbitrary $m,n$ (see 
also (\ref{x2})), it follows that
\begin{eqnarray}
W_1 & \leq &
\sum_{m,n}  \tr\{ \rho_nA P_m A\}
\nonumber
\\
& = &
\tr\left\{\left(\sum_n \rho_n\right)\, A \, \left(\sum_m P_m\right)\, A \right\} \ ,
\label{z11c}
\end{eqnarray}
where, similarly as in (\ref{1}) and (\ref{2}),
the sums run over the full range of 
admitted $m$ and $n$ values.
With (\ref{7}), (\ref{x0}), (\ref{z11a}) 
we thus obtain
\begin{eqnarray}
\Sigma_{1} & \leq & 
\underset{n\geq 2}{\textrm{max}}\,p_{n} 
\, 
\tr\{\bar\rho A^2\} \ .
\label{z12}
\end{eqnarray}

From the definition (\ref{a40}) it readily follows
that $\tilde A_{nm}=\tilde A_{mn}^\ast$.
Hence, $\Sigma_1$ can also be considered
as the partial sum on the right hand 
side of (\ref{:33A-1}) over all summands 
with $m=1$.
As a consequence, we can rewrite (\ref{:33A-1}) as
\begin{eqnarray}
S & \leq & g\ \Sigma_1 + g\  \Sigma'
\label{:39A}
\\
\Sigma' & := &
\sum_{m=2}^N \underset{\underset{n \not =m}{n=1}}{\sum^N}
|\tilde A_{mn}|^2
\ .
\label{:39A-1}
\end{eqnarray}
Analogously as in (\ref{z11a})-(\ref{z12}) one can conclude that
\begin{eqnarray}
\Sigma' & \leq & 
\underset{n\geq 2}{\textrm{max}}\,p_{n} 
\, 
\tr\{\bar\rho A^2\}
\ .
\label{z13}
\end{eqnarray}

For the sake of convenience only, we have
so far assumed that the largest energy level
population is given by $p_1$, see (\ref{x3a}).
In order to get rid of this convenient
but unnecessary special role of $n=1$,
we introduce (\ref{x3b}) into 
(\ref{z12}) and (\ref{z13}), 
yielding with (\ref{:39A}) 
\begin{eqnarray}
S & \leq & 2\, g\ 
{\max_{n}}' p_{n} 
\, 
\tr\{\bar\rho A^2\} 
\label{z14}
\end{eqnarray}
as the main result of the present subsection

%%%%%%%%%%%%%%%%%%%%%%%%%%%%%%%%%%%%%%%%%%%%%%%%
\subsubsection*{Final result} 
\label{a14}
We first address the case $\tr\{\bar\rho A^2\} =0$.
Since $A^2$ as well as all the summands on the right
hand side on (\ref{9}) are Hermitian, non-negative 
operators, it follows that
\begin{equation}
\tr\{P_n\rho(0)P_nA^2\} = 0
\label{y1}
\end{equation}
for all $n$.
Similarly as in (\ref{x4}) we can conclude  
from (\ref{3}) that
\begin{eqnarray}
| A_{mn}|^2 & = & \tr\{(P_m \rho^{1/2})(\rho^{1/2} P_n A) \}
\nonumber
\\
& \leq &
\tr\{ P_m \rho(0) P_m\}\, \tr\{ P_n \rho(0) P_n A^2 \} \ .
\label{y2}
\end{eqnarray}
With (\ref{y1}) it follows that $A_{mn}=0$ for all $m,n$
and with (\ref{2}) that $\tr\{\rho(t)A\}=0$ for all $t$.
Likewise, one finds with (\ref{9}) that
$\tr\{\bar \rho A\}=\sum_n A_{nn}=0$.
As a consequence, (\ref{1a}) is trivially 
fulfilled.

Next we turn to the case $\tr\{\bar\rho A^2\} >0$
(since $\bar\rho$ and $A^2$ are non-negative, the
case $\tr\{\bar\rho A^2\} <0$ is excluded).
It follows that
\begin{equation}
\beta:=g\  {\max_{n}}' p_{n} \, \tr\{\bar\rho A^2\} > 0 \ .
\label{y3}
\end{equation}
We thus can choose $d$ in (\ref{a23}) so 
that $\delta^2(t)\leq\beta/20$ for all $t$
and hence that
\begin{equation}
\left\langle \delta^{2}(t)\right\rangle_{T}
\leq\beta/20
\label{y4}
\end{equation}
for all $T>0$.
Likewise, we can choose $\epsilon=\beta/20$ in (\ref{:31A-3}), 
implying with (\ref{z14}) that
\begin{equation}
\left\langle \tilde \sigma^{2}(t)\right\rangle_{T}
\leq 2.05\, \beta
\label{y5}
\end{equation}
for all sufficiently large $T$.
In view of (\ref{:19A-2}) we can conclude that
\begin{eqnarray}
\left\langle \sigma^{2}(t)\right\rangle_{T}
& = &
\left\langle \tilde \sigma^{2}(t)\right\rangle_{T}
+2\, V
+\left\langle \delta^{2}(t)\right\rangle_{T}
\label{y6}
\\
V & := & \left\langle \tilde \sigma(t)\delta(t)\right\rangle_{T} \ .
\label{y7}
\end{eqnarray}
Observing that $\left\langle f_1(t) f_2(t)\right\rangle_{T}$
represents a well-defined scalar product for arbitrary
real valued functions $f_{1,2}(t)$,
the Cauchy-Schwarz inequality implies
\begin{eqnarray}
|V|^2\leq 
\left\langle \tilde \sigma^{2}(t)\right\rangle_{T}
\,\left\langle \delta^{2}(t)\right\rangle_{T} 
\ .
\label{y8}
\end{eqnarray}
With (\ref{y4}), (\ref{y5}) it follows that $|V|\leq 0.4\,\beta$,
and with (\ref{y6}) that
\begin{equation}
\left\langle \sigma^{2}(t)\right\rangle_{T}\leq 3 \, \beta
\label{y9}
\end{equation}
for all sufficiently large $T$.
Due to of (\ref{:26A}) and (\ref{y3}) we thus recover
(\ref{1a}).

%%%%%%%%%%%%%%%%%%%%%%%%%%%%%%%%%%%%%%%%%%%%%%%%
\section*{Appendix B}
\label{a2}
In this appendix we derive the general bound from 
Eq. (\ref{25}) by generalizing the approach of Short 
in Ref. \cite{sho11}.

We imagine many repetitions of the
measurement series considered in 
Sect. \ref{s4}.
As above Eq. (\ref{24}), in every repetition, 
either $\rho(t)$ or $\bar\rho$ is randomly 
selected with probability $1/2$ and then 
used to randomly generate a measurement 
outcome $\vec s$ according to the
corresponding probability (\ref{17}) 
or (\ref{18}).
But in contrast to Sect. \ref{s4},
the task is now to guess in every single 
repetition from the given data $\vec s$
whether $\rho(t)$ or $\bar\rho $ had 
been used to generate $\vec s$.

This decision problem is a generalization
of the one considered by Short \cite{sho11}
(see also Sect. \ref{s3b}).
It is in many respects also 
{\em similar} 
to those considered in Sect. \ref{s4}.
However, it is crucial to note that it is
{\em not identical} 
and that quantitative 
statements in one case do not immediately 
imply any rigorous conclusions in the 
other case (see also Sect. \ref{s3b}).
Yet, such rigorous conclusions are not 
impossible, as we will now show.

A key observation is that the above 
specified problem assigns well defined, 
objective probabilities (frequencies of 
occurrence) to each of the two ``models'' 
$\rho(t)$ and $\bar\rho$ 
(namely $1/2$ to each of them).
As a consequence, the conventional 
probabilistic (frequentist) approach 
happens to coincide with the concepts 
of Bayesian inference in this 
specific case.

In other words, in every single repetition
we are given the data $\vec s$ and
we have at our disposition the full 
knowledge about $\rho(t)$ and $\bar\rho$,
but about nothing else. 
Now we are forced to produce a 
decision based on this information.
The salient point consists in the observation 
that for any given $\vec s$ the only 
information of use is the pair of probabilities
$p_t(\vec s)$ and $\bar p(\vec s)$,
following from $\rho(t)$ and $\bar\rho$
according to (\ref{17a})-(\ref{18}).
Any other information contained in
$\rho(t)$, $\bar\rho$, and $\vec s$ 
is of no use for our decision problem.
Obviously (or by invoking Bayesian inference), 
the best one can do is to
opt for $\rho(t)$ if $p_t(\vec s)>\bar p(\vec s)$
and vice versa
(any other way of using the two 
numbers $p_t(\vec s)$ and $\bar p(\vec s)$ 
would not lead to a better decision).
In case $p_t(\vec s)=\bar p(\vec s )$ we introduce 
as a third option the answer ``undecided''
(alternatively, one could randomly choose
one of the two options with probability $1/2$).
Counting a correct decision as $1$ and a
wrong decision as $-1$, the success probability 
(symbol $\SP$),
%has nothing to do with 
%$S$ from (\ref{:30A-2})), 
i.e. the probability
of opting by means of the above optimal 
decision strategy for the correct 
state minus opting for the wrong 
state, follows as
\begin{eqnarray}
\SP_{opt} & = & \sum\limits_{\vec s}  \frac{p_t(\vec s)}{2}
\frac{1+\Do}{2}
\nonumber
\\
& + &
\sum\limits_{\vec s} \frac{\bar p(\vec s )}{2}
\frac{1-\Do}{2}
\label{1b}
\end{eqnarray}
\begin{eqnarray}
\Do & := & 1 \ \mbox{if $p_t(\vec s )>\bar p(\vec s )$}
\nonumber
\\
\Do & := & -1 \ \mbox{if $p_t(\vec s )<\bar p(\vec s )$}
\nonumber
\\
\Do & := & 0 \ \mbox{if $p_t(\vec s )=\bar p(\vec s )$} \ .
\label{2b}
\end{eqnarray}
The detailed justification is as follows:
The first factor, $p_t(\vec s)/2$, 
on the right hand side of (\ref{1b}) 
represents the joint probability that the 
random event $(\rho(t),\, \vec s)$ is realized.
The second factor, $[1+\Do]/2$ is unity if our 
guess was right, zero if it was wrong, and 
$1/2$ if we were undecided 
(or randomly picked one of the two options).
Similar considerations apply to the second 
sum in (\ref{1b}).

Since $\sum_{\vec s} p_t(\vec s )
=\sum_{\vec s} \bar p(\vec s )=1$, 
it readily follows from (\ref{1b}) that 
\begin{eqnarray}
\SP_{opt} & = & \frac{1}{2}+
\sum\limits_{\vec s} \frac{p_t(\vec s )-\bar p(\vec s )}{4} \, \Do
\label{3b}
\end{eqnarray}
Without loss of generality we can restrict the
sum to summands with $p_t(\vec s )\not =\bar p(\vec s )$
and rewrite $\Do$ from (\ref{2b})
for those summands as 
$|p_t(\vec s )-\bar p(\vec s )|/\{ p_t(\vec s )-\bar p(\vec s )\}$,
yielding
\begin{eqnarray}
\SP_{opt} & = & \frac{1}{2}+
\sum\limits_{\vec s} \frac{|p_t(\vec s )-\bar p(\vec s )|}{4} \ .
\label{4b}
\end{eqnarray}

Next, we consider the very same decision problem,
but now by employing any of the $\D $-functions 
from Sect. \ref{s4} as follows:
If $\D (\rho(t),\bar\rho ,\vec s) \geq 0$ then we opt
with probability $p_+:=\D (\rho(t),\bar\rho ,\vec s )$ 
for $\rho(t)$ and with
with probability $1-p_+$
our answer is ``undecided''
(randomly pick $\rho(t)$ or $\bar\rho $).
Likewise, if $\D (\rho(t),\bar\rho ,\vec s )<0$ then 
we opt with probability 
$p_-:=-\D (\rho(t),\bar\rho ,\vec s )$ 
for $\bar\rho $ and with probability
$1-p_-$ we are undecided. 
Similarly as in (\ref{1b}),
the success probability of this
decision strategy now takes the form
\begin{eqnarray}
\SP & = & \sum\limits_{\vec s} \frac{p_t(\vec s )}{2}
\frac{1+\D (\rho(t),\bar\rho ,\vec s )}{2}
\nonumber
\\
& + &
\sum\limits_{\vec s} \frac{\bar p(\vec s )}{2}
\frac{1-\D (\rho(t),\bar\rho ,\vec s )}{2}
\label{5b}
\end{eqnarray}
and like in (\ref{2b}) it follows that
\begin{eqnarray}
\SP & = & \frac{1}{2}+
\sum\limits_{\vec s} \frac{p_t(\vec s )-\bar p(\vec s )}{4} \, 
\D (\rho(t),\bar\rho ,\vec s ) \ .
\label{6b}
\end{eqnarray}
On the other hand, subtracting (\ref{24}) from (\ref{23}) yields
\begin{eqnarray}
\QQ (t)=\sum\limits_{\vec s} \frac{p_t(\vec s )-\bar p(\vec s )}{2} \D (\rho(t),\bar\rho ,\vec s ) \ .
\label{7b}
\end{eqnarray}
Upon comparison with (\ref{6b}) 
it follows that $\SP=[1+\QQ (t)]/2$. 
Since this success probability cannot 
exceed the optimal value $\SP_{opt}$ from (\ref{4b}),
we obtain $[1+\QQ (t)]/2\leq \SP_{opt}$. 
Likewise, by employing the decision strategy
$-\D (\rho(t),\bar\rho ,\vec s )$
instead of $\D (\rho(t),\bar\rho ,\vec s )$,
one recovers $[1-\QQ (t)]/2\leq \SP_{opt}$.
Combining both inequalities 
implies $[1+|\QQ (t)|]/2\leq \SP_{opt}$.
Together with (\ref{4b})
this yields our final result 
(\ref{25}).

Note that in order to derive this 
result we employed a different 
decision problem than the one considered
in Sect. \ref{s4}.
Yet, the so obtained inequality 
(\ref{25}) itself is valid 
independently of this specific 
decision problem.

%%%%%%%%%%%%%%%%%%%%%%%%%%%%%%%%%%%%%%%%%%%%%%%%
\section*{Appendix C}
\label{a3}
In this Appendix we provide the derivation of (\ref{26})-(\ref{27}).

Focusing on any of the projectors $\KK_\nu^{(j)}$
appearing in (\ref{17a}) and (\ref{17b}), one readily
finds upon replacing $A$ by $\KK_\nu^{(j)}$
in (\ref{11})-(\ref{14a})
that for any given $\epsilon_\nu^{(j)}>0$
\begin{eqnarray}
T_\nu^{(j)} / T \leq  
\alpha \, \tr\{\bar\rho\, [\KK_\nu^{(j)}]^2\} \, [\epsilon_\nu^{(j)}]^{-2} 
=
\alpha \, \bar \kk^{(j)}_\nu\, [\epsilon_\nu^{(j)}]^{-2}
\label{1c}
\end{eqnarray}
for all sufficiently large $T$,
where 
\begin{eqnarray}
\alpha & := & 
%24 
3 \, g\, {\max_n}' p_n
\label{2c}
\\
T_\nu^{(j)} & := & 
\big| 
\{\, 0 \leq t  \leq T\, : \, |\sigma_\nu^{(j)}(t)|  > \epsilon_\nu^{(j)}\, \}
\big| \ ,
\label{3c}
\\
\sigma_\nu^{(j)}(t) & := & 
\langle \KK_\nu^{(j)}\rangle_{\!\rho(t)} - \langle \KK_\nu^{(j)}\rangle_{\!\bar\rho}
= \kk^{(j)}_\nu (t) - \bar \kk^{(j)}_\nu
\  ,
\label{4c}
\end{eqnarray}
and where we exploited (\ref{17a}), (\ref{17b}),
and $[\KK_\nu^{(j)}]^2=\KK_\nu^{(j)}$
in the last equalities in (\ref{1c}) and (\ref{4c}).

For any given pair $(j,\nu)$
(where $j\in\{1,...,N_{rep}\}$ and
$\nu\in\{1,...,N_{A_j}\}$)
and any given $\epsilon_\nu^{(j)}>0$,
the quantity $T_\nu^{(j)}$ 
in (\ref{3c}) is the Lebesgue measure 
of all times $t\in [0, T]$ for which 
$|\sigma_\nu^{(j)}(t)|  > \epsilon_\nu^{(j)}$ 
holds true.
Since the number of pairs $(j,\nu)$ is finite,
it follows that for any given set of 
positive $\epsilon_\nu^{(j)}$ values
the inequality (\ref{1c}) applies simultaneously 
for all pairs $(j,\nu)$ provided $T$ 
is sufficiently large.
Hence, the measure of all times 
$t\in [0,T]$ for which 
$|\sigma_\nu^{(j)}(t)|  > \epsilon_\nu^{(j)}$ 
is true for at least one among all pairs 
$(j,\nu)$ can be estimated from above by
\begin{eqnarray}
T_{tot}:=\sum_{j=1}^{N_{rep}}\sum_{\nu=1}^{N_{A_j}} 
T_\nu^{(j)}
\label{5c}
\end{eqnarray}
for all sufficiently large $T$.
For all other times $t\in [0,T]$ it is true that
$|\sigma_\nu^{(j)}(t)|  \leq \epsilon_\nu^{(j)}$ 
simultaneously for all pairs $(j,\nu)$.
From now on, we exclusively focus 
on the latter subset of $[0,T]$, i.e. on times 
$t$ for which
\begin{eqnarray}
|\sigma_\nu^{(j)}(t)|  \leq \epsilon_\nu^{(j)}
\, \mbox{ for all $j\in\{1,...,N_{rep}\}$,
$\nu\in\{1,...,N_{A_j}\}$} \ .
\label{6c}
\end{eqnarray}
Thus, the Lebesgue measure of all times $t\in [0,T]$ 
for which the subsequently derived implications of
(\ref{6c}) may possibly not apply, is bounded 
by $T_{tot}$ from (\ref{5c}).

A particularly convenient choice of the
quantities $\epsilon_\nu^{(j)}$ turns out 
to be:
\begin{eqnarray}
\epsilon_\nu^{(j)} & := & \bar \kk_\nu^{(j)} \, F/N_{rep}\ 
\mbox{if $\bar \kk_\nu^{(j)} >\kk_{th}$}
\label{7c}
\\
\epsilon_\nu^{(j)} & := & (\kk_{th}\, \bar \kk_\nu^{(j)})^{1/2} \, F/N_{rep}\ 
\mbox{if $\kk_{th}\geq \bar \kk_\nu^{(j)} >0$}
\label{8c}
\\
\epsilon_\nu^{(j)} & := & \kk_{th} \, F/N_{rep}\ 
\mbox{if $\bar \kk_\nu^{(j)} = 0$} \ ,
\label{9c}
\end{eqnarray}
where the ``factor'' $F$ and the ``threshold''
$\kk_{th}$ are
positive real numbers, 
whose explicit values will be fixed later.
For the moment, we only require that
\begin{eqnarray}
0< F \leq 1/2 \ .
\label{10c}
\end{eqnarray}

Introducing (\ref{7c})-(\ref{9c}) into (\ref{1c}) 
and (\ref{5c}) implies for all sufficiently 
large $T$ that
\begin{eqnarray}
\frac{T_{tot}}{T} 
\leq
\sum_{j=1}^{N_{rep}}\sum_{\nu=1}^{N_{A_j}}
\alpha \frac{N_{rep}^2}{F^2\, \kk_{th}} \ .
\label{11c}
\end{eqnarray}
Exploiting (\ref{19}), the number of summands 
in the double sum can 
be readily bounded from above by 
$N_{rep}N_{obs}$,
%due to (\ref{19}), 
yielding
\begin{eqnarray}
\frac{T_{tot}}{T} 
\leq
\alpha \frac{N_{obs}\, N_{rep}^3}{F^2\, \kk_{th}}=:\epsilon \ .
\label{12c}
\end{eqnarray}
for all sufficiently large $T$.

Recalling the notation $\vec s:=(s_1,...,s_{N_{rep}})$
from below Eq. (\ref{17b}),
we divide the set of all possible 
measurement outcomes
\begin{eqnarray}
S := \{ \vec s 
%= (s_1,...,s_{N_{rep}}) 
\, : \, 
s_j \in \{1,...,N_{A_j}\}, \,
j\in\{1,...,N_{rep}\}\,  \}
\label{13c}
\end{eqnarray}
into the two subsets
\begin{eqnarray}
S' & := & \{ \vec s \in S 
\, : \, 
\bar \kk_{s_j}^{(j)} > \kk_{th} \, \mbox{for all $j$}\,  \}
\label{14c}
\\
S'' & := & S\setminus S' \ .
\label{15c}
\end{eqnarray}
Likewise, the sum over all $\vec s \in S$ 
appearing in (\ref{25}) is split into two parts 
according to
\begin{eqnarray}
\QQ _{max}(t) & = & (\Sigma' + \Sigma'')/2
\label{16c}
\\
\Sigma' & := & \sum\limits_{\vec s\in S'}
|p_t(\vec s ) - \bar p(\vec s )|  
\label{17c}
\\
\Sigma'' & := & \sum\limits_{\vec s\in S''}
|p_t(\vec s ) - \bar p(\vec s )|  \ .
\label{18c}
\end{eqnarray}

To evaluate $\Sigma'$, we note that
$\vec s\in S'$, 
implies $\bar \kk_{s_j}^{(j)} > \kk_{th}$
for all $j$ according to (\ref{14c}) and hence 
\begin{eqnarray}
|\sigma_{s_j}^{(j)}(t)| \leq  \bar \kk_{s_j}^{(j)} \, F/N_{rep}
\label{18c0}
\end{eqnarray}
according to (\ref{6c}) and (\ref{7c}).
With (\ref{10c}) and $N_{rep}\geq 1$
it follows that 
$|\sigma_{s_j}^{(j)}(t)| \leq \bar \kk_{s_j}^{(j)}/2$
and with (\ref{4c}) that 
\begin{eqnarray}
\kk_{s_j}^{(j)}(t) 
\geq 
\bar \kk_{s_j}^{(j)} - |\sigma_{s_j}^{(j)}| 
\geq \bar \kk_{s_j}^{(j)}/2
\label{18c1}
\end{eqnarray}
for all $j$.
Thus, all quantities in
Eqs. (\ref{17}) and (\ref{18}) 
are positive real numbers,
i.e., we can logarithmize those 
equations to obtain
\begin{eqnarray}
& & x:=\ln\left( \frac{p_t(\vec s)}{\bar p(\vec s)}\right)
= \sum_{j=1}^{N_{rep}} \ln\left(\frac{\kk^{(j)}_{s_j}(t)}{\bar \kk^{(j)}_{s_j}}\right)  
%> 0
\label{18c4}
\\
& & |p_t(\vec s) - \bar p(\vec s)| = |\bar p(\vec s) \, (e^x-1)|= \bar p(\vec s) \, |e^x-1| \ .
\label{18c3}
\end{eqnarray}

We first focus on the case $x\geq 0$.
Observing that $\ln(1+y)\leq y$ for all $y>-1$,
it follows that $\ln(a/b)=\ln(1+[a-b]/b)\leq [a-b]/b$ 
for all $a,\,b>0$, and hence with 
(\ref{18c4}), (\ref{4c}), and (\ref{18c0})
that
\begin{eqnarray}
0 \leq x \leq 
%{\sum_j}' 
\sum_{j=1}^{N_{rep}}
\frac{|\sigma^{(j)}_{s_j}(t)|}{\bar \kk^{(j)}_{s_j}}
\leq 
%{\sum_j}' 
\sum_{j=1}^{N_{rep}}
\frac{F}{N_{rep}} = F \ .
\label{18c5}
\end{eqnarray}
%where we exploited 
%(\ref{6c}) and (\ref{7c}) 
%(\ref{18c0}) in the second inequality.
%Observing %that $F\leq1/2$ according to 
%(\ref{10c}) %and that there are at most $N_{rep}$ 
%summands on the right hand side of (\ref{18c5})
%implies that $x\leq 1/2$ and 
In conclusion,
\begin{eqnarray}
|e^x-1| \leq  e^F-1
\label{18c6}
\end{eqnarray}
whenever $x\geq 0$ in (\ref{18c3}).
Turning to $x<0$, we observe that
\begin{eqnarray}
|e^x-1| & = & 
%1-e^{x}=
e^{x}(e^{-x}-1) <  e^{-x}-1 
\label{18c10}
\\
0 &  < & -x = \sum_{j=1}^{N_{rep}} \ln\left(\frac{\bar \kk^{(j)}_{s_j}}{\kk^{(j)}_{s_j}(t)}\right)  \ ,
\label{18c2}
\end{eqnarray}
where we exploited (\ref{18c4}) in the last step.
Similarly as in (\ref{18c5}) it follows that
\begin{eqnarray}
-x \leq \sum_{j=1}^{N_{rep}}
\frac{|\sigma^{(j)}_{s_j}(t)|}{\kk^{(j)}_{s_j}(t)}
\leq 
\sum_{j=1}^{N_{rep}}
\frac{\bar \kk^{(j)}_{s_j}}{\kk^{(j)}_{s_j}(t)} \frac{F}{N_{rep}} \ .
\label{18c8}
\end{eqnarray}
Since $\bar \kk_{s_j}^{(j)}/\kk_{s_j}^{(j)}(t) \leq 2$ 
according to (\ref{18c1}) we can conclude that
$-x \leq 2F$ and with (\ref{18c10}) that  $|e^x-1| < e^{2F}-1$
whenever $x < 0$ in (\ref{18c3}).
With (\ref{18c6}) we thus obtain
\begin{eqnarray}
|e^x-1| < e^{2F}-1
\label{18c9}
\end{eqnarray}
for arbitrary $x$ in (\ref{18c3}).
Due to the elementary inequality $e^z-1\leq (e-1)z\leq 2z$ 
for all $z\in [0,1]$ it follows with (\ref{10c}) 
that $e^{2F}-1\leq 4 F$ and hence with 
(\ref{18c3}) and (\ref{18c9}) that
\begin{eqnarray}
|p_t(\vec s) - \bar p(\vec s)|\leq  4 \, F \, \bar p(\vec s) \ .
\label{26c1}
\end{eqnarray}
Accordingly, $\Sigma'$ from (\ref{17c}) can 
be estimated as
\begin{eqnarray}
\Sigma' \leq 
4\, F\, 
%[e^{2F}-1] 
\sum_{\vec s\in S'} 
\bar p(\vec s)  
\leq 
4\, F\,
%[e^{2F}-1]
\sum_{\vec s\in S} \bar p(\vec s)=
4\, F \ .
%[e^{2F}-1] \  .
\label{27c}
\end{eqnarray}
%where we exploited that $\sum_{\vec s} \bar p(\vec s)=1$ 
%in the last step.

Next we upper bound $\Sigma''$ in (\ref{18c}) as
\begin{eqnarray}
\Sigma'' & \leq & \Sigma_1+\Sigma_2
\label{28c}
\\
\Sigma_1 & := & \sum\limits_{\vec s\in S''} p_t(\vec s )
\label{29c}
\\
\Sigma_2 & := & \sum\limits_{\vec s\in S''} \bar p(\vec s ) \ .
\label{30c}
\end{eqnarray}
Furthermore, we introduce the following subsets of
$S$ from (\ref{12c}):
\begin{eqnarray}
S_j & := & \{\vec s \in S \, : \, \bar \kk_{s_j}^{(j)}\leq \kk_{th}\} \ ,
\label{31c}
\end{eqnarray}
where $j=1,...,N_{rep}$.
According to (\ref{14c}) and (\ref{15c}) there exists
for every $\vec s\in S''$ at least one $j\in\{1,.,,,N_{rep}\}$
with the property that $\bar \kk_{s_j}^{(j)}\leq \kk_{th}$.
It follows that the union of all the subsets $S_j$
from (\ref{31c}) reproduce $S''$ and hence that 
\begin{eqnarray}
\Sigma_2 & \leq & \sum_{j=1}^{N_{rep}} \Sigma_2^{(j)}
\label{32c}
\\
\Sigma_2^{(j)} & := & 
\sum\limits_{\vec s\in S_j} \bar p(\vec s ) = 
\sum\limits_{\vec s\in S_j} \prod_{l=1}^{N_{rep}} \bar \kk^{(l)}_{s_l} \ ,
\label{33c}
\end{eqnarray}
where we exploited (\ref{18}) in the last step.
With (\ref{31c}) it follows that
\begin{eqnarray}
\Sigma_2^{(j)} 
\leq  
\sum\limits_{\vec s\in S_j} \kk_{th}
\prod_{l\not=j}^{N_{rep}} \bar \kk^{(l)}_{s_l} 
\leq  
\sum\limits_{\vec s\in S} \kk_{th}
\prod_{l\not=j}^{N_{rep}} \bar \kk^{(l)}_{s_l} 
\ ,
\label{34c}
\end{eqnarray}
where the symbol $l\not=j$ indicates 
that the $j$-th factor is omitted and where
we exploited that $S_j\subset S$ 
in the last step (cf. (\ref{31c})).
In view of (\ref{13c}) we can 
conclude that
\begin{eqnarray}
\Sigma_2^{(j)} 
\leq  
\sum_{s_j=1}^{N_{A_j}} \kk_{th}
\left(\prod_{l\not=j}^{N_{rep}} \sum_{s_l=1}^{N_{A_l}} 
\bar \kk^{(l)}_{s_l} \right) = \kk_{th}\,N_{A_j} 
\label{35c}
\end{eqnarray}
where we exploited 
that $\sum_{s_l=1}^{N_{A_l}} \bar \kk^{(l)}_{s_l}=1$ 
for all $l$.
Taking into account (\ref{19}) and (\ref{32c})
we finally obtain
\begin{eqnarray}
\Sigma_2 \leq  \kk_{th}\, N_{obs}\, N_{rep} \ .
\label{36c}
\end{eqnarray}

Similarly as in (\ref{32c})-(\ref{34c}) it follows 
with (\ref{29c}) and (\ref{17}) that
\begin{eqnarray}
\Sigma_1 & \leq & \sum_{j=1}^{N_{rep}} \Sigma_1^{(j)}
\label{37c}
\\
\Sigma_1^{(j)} & := & 
\sum\limits_{\vec s\in S_j} p_t(\vec s ) = 
\sum\limits_{\vec s\in S_j} \prod_{l=1}^{N_{rep}} \kk^{(l)}_{s_l} (t)\ ,
\nonumber
\\
& \leq & 
\sum\limits_{\vec s\in S_j} \kk^{(j)}_{s_j} (t)
\prod_{l\not=j}^{N_{rep}} \kk^{(l)}_{s_l} (t) \ .
\label{38c}
\end{eqnarray}
For all $s_j$ appearing in the last sum over
$\vec s\in S_j$ Eq. (\ref{31c}) implies that
$\bar \kk_{s_j}^{(j)}\leq \kk_{th}$
and hence that $\epsilon_{s_j}^{(j)} \leq \kk_{th}$
according to (\ref{8c})-(\ref{10c}).
With (\ref{4c}) and (\ref{6c})
we thus can infer that 
\begin{eqnarray}
\kk^{(j)}_{s_j} (t)
\leq 
\bar \kk^{(j)}_{s_j} + |\sigma^{(j)}_{s_j} (t)|
\leq 
\bar \kk^{(j)}_{s_j} + \epsilon_{s_j}^{(j)} 
\leq 2 \, \kk_{th} \ .
\label{39c}
\end{eqnarray}
By combining this result with (\ref{38c}) 
one finds exactly as in
(\ref{34c})-(\ref{35c}) that
\begin{eqnarray}
\Sigma_1^{(j)} 
\leq  
2 \sum\limits_{\vec s\in S_j} \kk_{th}
\prod_{l\not=j}^{N_{rep}} \kk^{(l)}_{s_l} (t)
\leq
2\, \kk_{th}\,N_{A_j} \ .
\label{40c}
\end{eqnarray}
Like in (\ref{36c}) it follows that
$\Sigma_1 \leq 2\, \kk_{th}\, N_{obs}\, N_{rep}$
and with (\ref{28c}) that
\begin{eqnarray}
\Sigma'' \leq 3\, \kk_{th}\, N_{obs}\, N_{rep} \ .
\label{41c}
\end{eqnarray}

Introducing (\ref{27c}) and (\ref{41c}) 
into (\ref{16c}) implies
\begin{eqnarray}
Q_{max}(t) \leq
%[e^{2F}-1 +3G]/2
[4F + 3 \kk_{th}N_{obs}N_{rep}]/2  \ ,
\label{42c}
%\\
%G & := & \kk_{th}\, N_{obs}\, N_{rep} \ ,
%\label{43c}
\end{eqnarray}
where $\kk_{th}>0$ and $F\in(0,1/2]$ can
still be chosen arbitrarily (see below (\ref{9c})).
We thus may choose $k_{th}$ so that
the right hand side of (\ref{42c}) equals 
$\epsilon$ from (\ref{12c}),
i.e.,
\begin{eqnarray}
k_{th} & = & [2\epsilon-4F]/3N_{obs}\, N_{rep} \ .
\label{48c}
\end{eqnarray}
%$3G=4F$, i.e. $\kk_{th}=4F/3N_{obs}\, N_{rep}$.
%$3G=e^{2F}-1$, i.e. $\kk_{th}=[e^{2F}-1]/3N_{obs}\, N_{rep}$.
Altogether, Eqs. (\ref{12c}), (\ref{42c}), and (\ref{48c}) imply
for all sufficiently large $T$ that
\begin{eqnarray}
T_{tot}/T & \leq & \epsilon
\label{45c}
\\
Q_{max}(t) & \leq & \epsilon
\label{46c}
\\
\epsilon & = & \frac{3 \alpha N_{obs}^2\, N_{rep}^4}{2 F^2[\epsilon - 2F]}
\label{47c}
\end{eqnarray}
Finally, we make the choice $F=\epsilon/3$, which is 
obtained by minimizing (\ref{47c}) with respect to $F$.
Upon inserting $\alpha$ from (\ref{2c}) and $F=\epsilon/3$
into (\ref{47c}) and then solving for $\epsilon$ 
%yields
%\begin{eqnarray}
%\epsilon & := & 3 \, (12\, g\, {\max_n}' p_n)^{1/4}\, N_{obs}^{1/2}\, N_{rep}
%\label{50c}
%\end{eqnarray}
%Observing that $12^{1/4}=1.86... <2$ 
one recovers (\ref{27}).

As announced below (\ref{6c}),
the result (\ref{46c}) is valid for all $t \in[0,T]$
apart from a subset of $[0,T]$, whose Lebesgue 
measure is bounded by $T_{tot}$, 
and provided $T$ is sufficiently large.
It follows that $T^\ast$ from (\ref{28}) 
cannot exceed $T_{tot}$, 
i.e. $T^\ast\leq T_{tot}$.
Upon comparison with (\ref{45c}), 
we thus recover (\ref{26}).

Strictly speaking, the above
conclusions are only valid if our
choice $F=\epsilon/3$
(see below (\ref{47c}))
is self-consistent with the 
constraint from (\ref{10c}).
Equivalently, this means that 
$\epsilon$ must be smaller 
than $3/2$.
In the opposite case, i.e.,
if $\epsilon$ in (\ref{27})
should happen to exceed 
$3/2$, then our above arguments
no longer apply, but obviously
(\ref{26}) is still trivially fulfilled.

%%%%%%%%%%%%%%%%%%%%%%%%%%%%%%%%%%%%%%%%%%%%%%%%

\end{document}